\documentclass[aps,pra,reprint,superscriptaddress,floatfix]{revtex4-2}

\usepackage[colorlinks=true,urlcolor=blue,citecolor=magenta]{hyperref}

\usepackage{amsmath,amssymb,bm}
\usepackage{graphicx}

\usepackage{dcolumn}
\usepackage[dvipsnames]{xcolor}

\begin{document}

\title{Suppression of capillary instability in a confined quantum liquid filament}%

\author{Francesco Ancilotto}\email{francesco.ancilotto@unipd.it}
\affiliation{\mbox{Dipartimento di Fisica e Astronomia `Galileo Galilei' and
CNISM, Universit\`{a} di Padova, 35131 Padova, Italy}}
\affiliation{\mbox{CNR-Officina dei Materiali (IOM), via Bonomea, 265 - 34136 Trieste, Italy}}

\author{Michele Modugno}
\affiliation{Department of Physics, University of the Basque Country UPV/EHU, 48080 Bilbao, Spain}
\affiliation{IKERBASQUE, Basque Foundation for Science, 48013 Bilbao, Spain}
\affiliation{EHU Quantum Center, University of the Basque Country UPV/EHU, Leioa, Biscay, Spain}

\author{Chiara Fort}\email{chiara.fort@unifi.it}
\affiliation{\mbox{Dipartimento di Fisica e Astronomia, Universit\`{a}
degli Studi di Firenze, 50019 Sesto Fiorentino, Italy}}
\affiliation{\mbox{European Laboratory for Non-Linear Spectroscopy, Universit\`{a}
degli Studi di Firenze, 50019 Sesto Fiorentino, Italy}}
\affiliation{\mbox{Istituto Nazionale di Ottica, CNR-INO, 50019 Sesto Fiorentino, Italy}}

\date{\today}

\begin{abstract}
Quantum Bose-Bose mixtures with strong attraction can form self-bound, liquid-like droplets stabilized by quantum fluctuations. Despite equilibrium densities much lower than those of classical liquids, these droplets exhibit finite surface tension and liquid-like behaviors. Recent experiments have demonstrated Rayleigh--Plateau instability in elongated droplets confined in an optical waveguide. 
Here we consider the case of an infinite filament and extend the theoretical description to include transverse harmonic confinement. By solving the Bogoliubov—de Gennes equations within a single-component framework, benchmarked against full Gross--Pitaevskii simulations, we show that increasing confinement progressively suppresses the instability, leading to complete stabilization beyond a critical trap frequency.

\end{abstract}

\maketitle


\section{\label{introduction}Introduction }
In recent years, various experimental and theoretical studies have investigated the physics of quantum Bose-Bose mixtures in the strongly attractive regime, where the competition between the mean-field (MF) attractive interaction and the repulsive Lee--Huang--Yang (LHY) contribution --accounting for quantum fluctuations-- leads to the formation of a liquid-like, self-bound state known as a quantum droplet.
This unique system was theoretically predicted by D.S.~Petrov in 2015 \cite{Petrov2015} and later observed experimentally in dipolar gases \cite{Kadau2016,FerrierBarbut2016,Schmitt2016,Ferlaino2016,FerrierBarbutJPB2016,Wenzel2017} as well as in homonuclear \cite{Cabrera2018,Semeghini2018} and heteronuclear \cite{Derrico2019,Burchianti2020,Guo2021}  mixtures. 
Quantum droplets are very peculiar self-bound states characterized by ultralow equilibrium densities and finite surface tension. Their liquid-like behavior enables the emergence of phenomena typical of classical liquids, thereby opening new avenues for exploring hydrodynamic instabilities in quantum many-body systems.

A compelling development in this direction is the recent observation of the capillary (or Rayleigh-Plateau) instability in an elongated quantum droplet confined within an optical waveguide \cite{Cavicchioli2025}.
This surface tension-driven instability is a hallmark of classical liquid filaments, leading to their breakup into droplets. It was originally predicted for an infinite cylinder of inviscid, incompressible liquid \cite{Plateau1857,Rayleigh1878,Rayleigh1879}. This phenomenon has been extensively studied, taking into account viscosity \cite{Rayleigh1892,Castrejon_PRL2012}, the effect of a different surrounding liquid, and boundary conditions \cite{Tomotika1935,Goren1962,Hammond1983, Zhao2023}. It has also been invoked to explain nuclear scission \cite{Brosa1990} and observed in multi-material optical fibers \cite{Shabahang2011}.
For an extended review of the subject see, e.g., Ref.~\cite{Eggers_2008}. In certain contexts, this instability is undesirable, and various strategies have been developed or proposed to suppress it (see Ref.~\cite{Patankar2022,Patrascu2019} and references therein).

The Rayleigh--Plateau instability has also been observed \cite{Speirs2020} and theoretically investigated in quantum liquids such as superfluid helium (see Ref.~\cite{Ancilotto2023b}). In addition, it has been studied in quantum filaments formed in immiscible Bose-Bose mixtures, where a cylinder of one component is surrounded by another component, leading to interfacial tension due to repulsive interspecies interactions \cite{Sasaki2011}, as well as
in attractive self-bound mixtures in free space \cite{Ancilotto2023}.

Here, we extend the treatment presented in Ref.~\cite{Ancilotto2023} to compute the full excitation spectrum of unstable modes, also including the effect of a waveguide, which introduces a transverse confinement to the filament. Throughout this work, we use the term \textit{filament} to denote an infinitely extended cylinder.
In particular, we study the spectrum of unstable modes in a Bose-Bose liquid-like filament by solving the Bogoliubov--de Gennes (BdG) equations using an effective single-component energy density functional that describes the mixture within a mean-field approach that includes quantum fluctuation effects via a LHY correction. 
The validity of this approximation is assessed by comparing its predictions with numerical simulations based on two coupled extended Gross--Pitaevskii (GP) equations.
We compare the calculated spectra with the Rayleigh--Plateau expression for the capillary instability of a classical inviscid,  incompressible liquid, extracting the critical parameters that characterize this instability, namely, the filament's radial size and the capillary time.

We find that harmonic transverse confinement stabilizes the filament, progressively suppressing the instability as the confinement strength increases. 
For sufficiently strong confinement, the filament becomes fully stable. 
This stabilization mechanism is interpreted in light of the hydrodynamic instability of classical inviscid fluids subjected to radial forces \cite{Patankar2022}, thus highlighting yet another connection with hydrodynamic phenomena reported in ultra-cold quantum gases that mirror the behavior of classical fluids \cite{Hernandez-Rajkov2024, Huh2024, Geng2024}.

This paper is organized as follows. 
In Sec.~\ref{sec:system}, we introduce the system under investigation and define the parameters used throughout this work.
In Sec.~\ref{sec:model}, we present the theoretical framework, beginning with the extended GP theory for a binary mixture, which includes the LHY correction beyond the mean field approximation (Sec.~\ref{sec:generalizedGP}), followed by the single-component approximation (Sec.~\ref{sec:single-comp}), and the corresponding BdG equations (Sec.~\ref{sec:BdG}).
In Sec.~\ref{sec:freespace}, we perform the stability analysis of a filament in free space, comparing the BdG spectrum of unstable modes (Sec.~\ref{sec:BdGanalysis}) with full numerical simulations based on coupled extended GP equations (Sec.~\ref{sec:GPcomparison}). By fitting the spectrum of the unstable modes with the Rayleigh--Plateau formula for the an inviscid liquid filament (Sec.~\ref{sec:RPinterpretation}), we extract the filament's radial size and capillary time and compare them with corresponding estimates obtained directly from the filament density distribution. The good agreement obtained confirms our interpretation in terms of capillary instability.
Finally, in Sec.~\ref{sec:confinedfilament}, we analyze the effect of an external radial harmonic confinement, finding a critical frequency above which the filament becomes stable. 
Conclusions are drawn in Sec.~\ref{sec_conclusions}.

\section{System}
\label{sec:system}

\begin{figure}[t]
\centerline{\includegraphics[width=0.9\linewidth,clip]{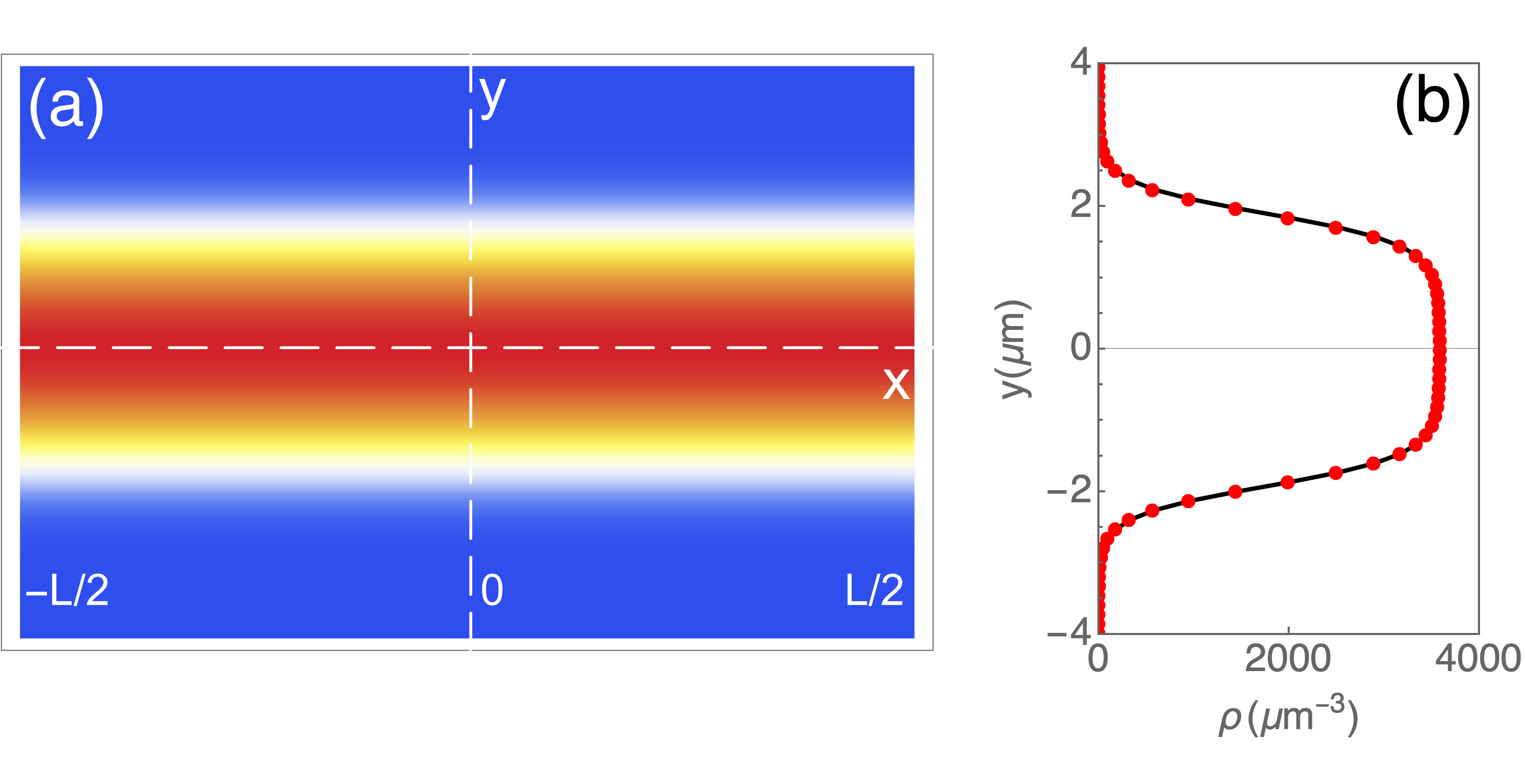}}
\caption{(a) Schematic representation of the filament and the system geometry (see text). The plot shows the calculated total density profile integrated along the transverse $z$ direction for a linear density of $N/L=4.2\times 10^{4} \mu \mathrm{m}^{-1}$.
(b) Corresponding radial profile obtained from the full two-component GP functional [Eq.~\eqref{eq:energy2c}] (points) and from the single-component approach of Sec.~\ref{sec:single-comp} (line).}
\label{fig:fig1}
\end{figure}

We consider a $^{41}\mathrm{K}$--$^{87}\mathrm{Rb}$ heteronuclear bosonic mixture in the self-bound droplet regime, forming a filament \footnote{The infinitely extented filament is a stationary state of the system and is therefore subject to instability. The true ground state would, instead, correspond to a spherical droplet in free space or to an elongated droplet in a radial waveguide}, elongated and homogeneous along the $x$ axis, as illustrated in Fig.~\ref{fig:fig1}(a)  (to perform numerical calculations, we consider a finite-size system of axial length $L$ with periodic boundary conditions).

The two species are labeled as $i=1$ for $^{41}\mathrm{K}$ and $i=2$ for $^{87}\mathrm{Rb}$.
The scattering lengths characterizing the intraspecies interactions are positive and fixed at $a_{11} = 62\,a_0$ and $a_{22} = 100.4\,a_0$, where $a_0$ is the Bohr radius. The heteronuclear scattering length $a_{12}$, which can be experimentally tuned by means of Feshbach resonances, is also fixed throughout this work at $a_{12} = -90\,a_0$ to ensure that the system lies well within the self-bound regime. 
This value is indeed beyond the onset of the mean-field collapse, leading to the quantum liquid state, which for the present mixture occurs at the critical value $a_{12} = -73.6\,a_0$. 
This critical value corresponds to the condition $g_{12} + \sqrt{g_{11}g_{22}} = 0$, where the coupling constants $g_{ij}$ are defined as $g_{ii}=4\pi\hbar^{2}a_{ii}/m_{i}$ ($i=1,2$), and $g_{12}=g_{21}=2\pi\hbar^{2}a_{12}/m_r$, with $m_i$ the mass of the $i$ component, and $m_r =m_1m_2/(m_1+m_2)$ the reduced mass.

In the bulk of the filament, the density ratio $\eta \equiv \rho_1/\rho_2$
is locked to the value $\eta =\sqrt{g_{22}/g_{11}}$,            
corresponding to the equilibrium ratio of a homogeneous mixture \cite{Petrov2015}. The system is then characterized by the total linear density, $N/L$, where $N=N_{1} + N_{2}$, with $N_{i}$ denoting the number of atoms of species $i$.
In the following, we refer to three specific values of $N/L$, taken as representative examples of filaments with a \textit{flat-top} radial profile ($N/L = 20.4 \times 10^4\;\mu\mathrm{m}^{-1}$), an intermediate case ($N/L = 4.2 \times 10^4\;\mu\mathrm{m}^{-1}$), and a low-density case ($N/L = 2.1 \times 10^4\;\mu\mathrm{m}^{-1}$), where no clear bulk region can be identified.
See Fig.~\ref{fig:fig2}, where the radial profiles of the filaments are shown.

In this paper, we analyze two distinct cases: (A) a filament in free space, and (B) a filament subjected to transverse harmonic confinement. In the latter case, we assume that both components experience the same external potential, 
\begin{equation}
V^{\mathrm{ext}}( r_\perp) = \frac{1}{2} m_1\Omega_1^2r_{\perp}^2 = \frac{1}{2} m_2\Omega_2^2r_{\perp}^2,
\label{eq:potential}
\end{equation}
where $r_\perp=\sqrt{y^2+z^2}$ represents the transverse (radial) coordinate.
In the following, we refer to the strengths of the transverse confinement by specifying the trap angular frequency $\Omega_1$, which we denote simply as $\Omega$, omitting the index for simplicity.

\section{Theoretical framework}
\label{sec:model}

In this section, we present the GP framework for describing a heteronuclear bosonic mixture in the self-bound droplet regime, and derive the single-component energy functional.

\subsection{Generalized Gross--Pitaevskii theory}
\label{sec:generalizedGP}

The generalized GP energy functional for an inhomogeneous mixture, including the usual mean-field terms and the LHY quantum correction, is \cite{Petrov2015,Ancilotto2018}
\begin{widetext}
\begin{equation}
E[\psi_{i}] = 
\sum_{i=1}^{2}\int\!\left[\frac{\hbar^2}{2m_i}|\nabla \psi_{i}(\bm{r})|^2 
+ V_i^{\mathrm{ext}}(\bm{r}) \rho_i(\bm{r})\right]d\bm{r} 
+\frac{1}{2}\sum_{i,j=1}^{2}g_{ij}\int\!\rho_{i}(\bm{r})\rho_{j}(\bm{r}) d\bm{r} 
+\int\!{\cal E} _{LHY}(\rho_1(\bm{r}),\rho_2(\bm{r}))d\bm{r},
\label{eq:energy2c}
\end{equation}
\end{widetext}
where, $\rho_i(\bm{r})=|\psi _i(\bm{r})|^2$ represent the number density of each component, normalized such that $\int  
\rho_{i}(\bm{r})d\bm{r} = N_i$.
The LHY term in the energy functional $E[\psi_{i}]$ is given by 
\begin{align}
{\cal E} _{\rm LHY} &= \frac{8}{15 \pi^2} \left(\frac{m_1}{\hbar^2}\right)^{3/2}
\!\!\!\!\!\!(g_{11} \rho_1)^{5/2}
f\left(\frac{m_2}{m_1},\frac{g_{12}^2}{g_{11}g_{22}},\frac{g_{22}\,\rho_2}{g_{11}\,\rho_1}\right),
\label{functional}
\end{align}
where $f$ is a dimensionless function, typically approximated by its value at  $g_{12}^2/(g_{11}g_{22})=1$ \cite{Petrov2015}. 
For the case of different masses, its explicit expression has been obtained in Ref.~\cite{Ancilotto2018}.
For numerical implementation, it can be conveniently approximated by a fitting function \cite{Minardi2019}. 
In our calculations, we use the following fit,
\begin{equation}
   f\left(\frac{m_2}{m_1},1,\xi\right) =1+a_f \xi+b_f \xi^2+c_f \xi^3+d_f \xi^{3/2}+e_f \xi^{5/2},
   \label{eq:fit}
\end{equation}
where we have defined $\xi\equiv g_{22}\,\rho_2/(g_{11}\,\rho_1)$.
The corresponding fit coefficients are given in Table~\ref{table}.
\begin{table}[h]
\begin{tabular}{cccccccccc}
\hline
& $a_f$  & $b_f$ & $c_f$  & $d_f$  & $e_f$ \\
\hline
 & 3.34520  & 1.81955  & 0.06753  & 1.36488  &  2.54208 \\
 \end{tabular}    
  \caption{Coefficients of the fitting function $f({m_2}/{m_1},1,\xi)$ in Eq.~\eqref{eq:fit}, for $m_2/m_1=87/41$.}
\label{table}
\end{table}

The time evolution of this system is governed by two coupled generalized GP equations, derived from the energy functional \eqref{eq:energy2c} via a variational principle, $i\hbar\partial_t \psi_i = \delta E / \delta \psi_i^*$ \cite{dalfovo1999}. This yields
\begin{equation}
  i \hbar \frac{\partial \psi _i}{\partial t} =
  \left[-\frac{\hbar^2}{2m_i}\nabla ^2 + V_i + \mu_{i}(n_1,n_2) \right]\psi _i
   \, , 
  \label{eq:gpe}
\end{equation}
with \cite{Derrico2019}  
\begin{equation}
  \mu_{i}\equiv \frac{\delta E}{\delta n_{i}} = g_{ii}n_i+g_{ij}n_j+
  \frac{\partial {\cal E}_{\rm LHY}}{\partial n_i}\quad \,( j \ne i) \,.
  \label{eq:chempot}
\end{equation}

\subsection{Single-component energy functional}
\label{sec:single-comp}

Assuming everywhere a fixed density ratio, $\eta \equiv \rho_1/\rho_2 = \sqrt{g_{22}/g_{11}}$ (see Sec.~\ref{sec:system}), the functional $E[\psi_{i}]$ can be reduced to an effective single-component form \cite{Jorgensen2018,Minardi2019,Cikojevic2021,Ancilotto2023}.

Let us define the following coefficients:
\begin{align}
        \label{alpha}
        \alpha ^\prime &= \frac{\hbar^2}{2m_1}+\frac{\hbar^2}{2m_2}\left(\frac{1}{\eta}\right),
        \\
        \label{beta}
        \beta ^\prime &= \frac {1}{2}g_{11}+ \frac {1}{2}g_{22}\left(\frac{1}{\eta^2}\right)+g_{12}\left(\frac{1}{\eta}\right),
        \\
        \label{gamma}
        \gamma ^\prime &= \frac{8}{15 \pi^2}\left(\frac{m_1}{\hbar^2}\right)^{3/2}g_{11}^{5/2} 
     f\left(\frac {m_2}{m_1},1,\sqrt{\frac {g_{22}}{g_{11}}}\right).
\end{align}
Then, the effective single-component energy density of the mixture, expressed in terms of the total density $\rho=\rho_1 + \rho_2$, reads
\begin{equation} 
\mathcal{E} = \alpha |\nabla \Psi(\bm{r}) |^2  + \beta \rho(\bm{r}) ^2 + \gamma \rho(\bm{r}) ^{5/2} + V^{\mathrm{ext}}(\bm{r})\rho(\bm{r}),
\label{eq:en_dens}
\end{equation}
where $|\Psi(\bm{r})|^2=\rho$ and
\begin{equation}
\alpha = \alpha ^\prime \left(\frac {\eta}{1+\eta}\right),\,
\beta = \beta ^\prime \left(\frac {\eta}{1+\eta}\right)^2, \,
\gamma=\gamma^\prime \left(\frac {\eta}{1+\eta}\right)^{5/2}. 
\end{equation}
In writing Eq.~\eqref{eq:en_dens}, we have made use of the fact that both components are subjected to the same external potential (see Sec.~\ref{sec:system}).

The minimization of the above energy functional
leads to the following Euler--Lagrange equation:
\begin{equation}
 H \Psi _0(\bm{r})=\mu  \Psi_0(\bm{r}),
 \label{eq:acca}
\end{equation}
where
\begin{equation}
 H = -\alpha \nabla^{2} + 2\beta \rho(\bm{r}) + \frac{5}{2} \gamma \rho^{3/2}(\bm{r}) + V^{\mathrm{ext}}(\bm{r}),
 \label{eq:nlgpe}
\end{equation}
and $\mu $ is a Lagrange multiplier determined by the normalization condition $\int |\Psi_0(\bm{r})|^2 d\bm{r}=N$.

The single-component functional introduced above accurately captures the behavior of the binary mixture in the self-bound regime, even in the presence of significant density inhomogeneities and deviations from the equilibrium bulk densities. 
For instance, Fig.~\ref{fig:fig1}(b) shows the radial density profile of a filament with $N/L=4.2\times 10^{4} \mu \mathrm{m}^{-1}$, comparing the prediction of the single-component approach (solid line) with the results obtained from the full two-component functional in Eq.~\eqref{eq:energy2c} (data points). 
The two methods exhibit excellent agreement.

\subsection{Bogoliubov--de Gennes equations}
\label{sec:BdG}

In this section, we compute the excitation spectrum $\omega (\bm{k})$ of the filament, within the BdG approximation, focusing on the onset of dynamical instability, identified by modes with $\omega^2(\bm{k}) < 0$. 
As anticipated, we consider two distinct cases: (A) a filament in free space, and (B) a filament subjected to a transverse harmonic confinement (see Sec.~\ref{sec:system} for details).

As outlined in the Appendix~\ref{sec:appendix}, the BdG excitation frequencies $\omega (\bm{k})$ are obtained by solving the non-Hermitian eigenvalue problem
\begin{equation}\everymath{\displaystyle}
\begin{bmatrix}
{A} & {B}  \\
-{B}  & -{A}  \\
\end{bmatrix}\begin{pmatrix} {u} \\ {v}
\end{pmatrix}=\hbar \omega
\begin{pmatrix} {u}
\\ {v}
\end{pmatrix},
\label{eig_AB}
\end{equation}
where 
\begin{align}
{A}_{\bm{G},\bm{G}^\prime}
&\equiv \delta _{\bm{G},\bm{G}^\prime}[{\alpha
}(\bm{k}+\bm{G})^2 -\mu 
] + V^{\mathrm{ext}}_{\bm{G}-\bm{G}^ \prime}
\nonumber \\
&\quad 
+ 2\tilde {U}^{MF}_{\bm{G}-\bm{G}^ \prime} 
+\frac {5}{2}\tilde {U}^{QF}_{\bm{G}-\bm{G}^ \prime}
\label{mat_A}
\\
{B}_{\bm{G},\bm{G}^\prime}
&\equiv -\tilde {U}^{MF}_{\bm{G}-\bm{G}^ \prime} 
-\frac {3}{2} \tilde {U}^{QF}_{\bm{G}-\bm{G}^ \prime}
\label{mat_B},
\end{align}
and the $\bm{G}$-vectors are the reciprocal lattice vectors associated with the periodic computational box (see Appendix~\ref{sec:appendix}). 
Note that the eigenvalue problem in Eq.~\eqref{eig_AB} can be reduced to a non-Hermitian problem of {\it half} the dimension (thus largely reducing the computational cost of diagonalization) by means of a unitary transformation:
\begin{equation}
(A-B) (A+B) |u+v\rangle = (\hbar \omega )^2|u+v\rangle.
\label{unitary}
\end{equation}

\section{
filament in free space}
\label{sec:freespace}

We start by considering the problem in free space ($V^{\mathrm{ext}} \equiv 0$), complementing the analysis of Ref.~\cite{Ancilotto2023} with a direct calculation of the growth rates of the unstable modes using the BdG formalism discussed above. These results, obtained with the effective single-component functional, will then be compared with those from the GP simulation of the full two-component system.

\subsection{BdG analysis}
\label{sec:BdGanalysis}

\begin{figure}
\centerline{
\includegraphics[width=0.8\linewidth,clip]{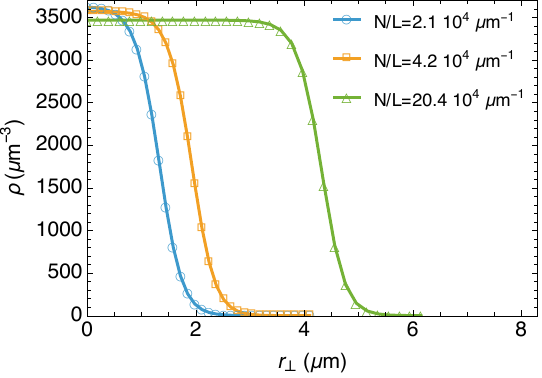}
}
\caption{Radial density profiles of the filament, for the three representative values of the linear density considered in this work (see legend).}
\label{fig:fig2}
\end{figure}

The stationary state of the filament is obtained by solving Eq.~\eqref{eq:acca}.
The corresponding density profiles along the radial direction are shown in Fig.~\ref{fig:fig2}, for the three different values of the linear density $N/L$ introduced in Sec. \ref{sec:system}.
We solve the eigenvalue problem \eqref{unitary} to obtain the excitation frequencies of modes polarized along the filament axis, i.e. $\omega (\bm{k})$ with $\bm{k}=(k,0,0)$.

\begin{figure}
\centerline{
\includegraphics[width=0.9\linewidth,clip]{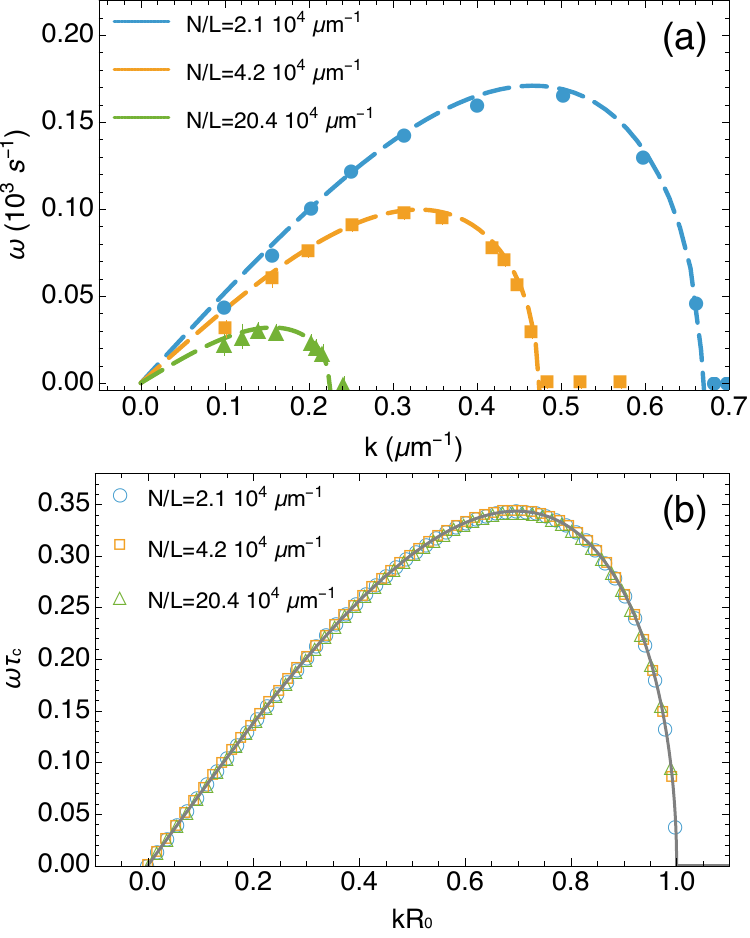}}
\caption{(a) Dashed lines: imaginary part of the lowest eigenvalue of Eq.~\eqref{eig_AB} as a function of the axial wavevector $k$, for the three different values of the linear density (see legend). 
Data points represent the corresponding spectra obtained from numerical solutions of the GP equations, as described in the text.
(b) Rescaled BdG spectra (see text); the solid line shows the classical prediction given by Eq.~\eqref{omkCF}.
}
\label{fig:fig3}
\end{figure}

The stability of a liquid (inviscid and incompressible) filament of radius $R$ is known to undergo capillary instability for any axial perturbation with wavevector $k < 1/R$. 
This instability
ultimately leads to the fragmentation of the filament into droplets.
In this context, imaginary excitation frequencies signal a dynamically unstable regime, where perturbations grow exponentially and drive the filament breakup. Indeed, we find that the lowest eigenvalue of Eq.~\eqref{unitary} corresponds to an imaginary frequency, $\omega ^2<0$, for wavevectors $k<k_c$, signaling the onset of capillary instability. 
The critical wavevector $k_c$ depends on the linear density $N/L$, as shown in Fig.~\ref{fig:fig3}(a), which displays the unstable modes spectra for the three representative values of the linear density considered.

\subsection{Comparison with GP simulations}
\label{sec:GPcomparison}

The growth rate of the unstable modes of the filament can also be obtained from the numerical solution of the coupled GP equations introduced in Sec.~\ref{sec:generalizedGP}.
We compute the initial stationary configurations by numerically minimizing the energy functional in Eq.~\eqref{eq:energy2c} through imaginary-time evolution \footnote{Stationary configurations of the energy functional \eqref{functional} are obtained using a steepest descent algorithm \cite{press2007}, which iteratively propagates the two condensate components in imaginary time. A cylindrical filament--rather than a spherical droplet--can be obtained by choosing suitable initial conditions. 
}. 

Then, to trigger the capillary instability, we add to the stationary solutions $\psi_i^{\mathrm{st}}$ ($i = 1, 2$) an axially symmetric perturbation $\delta\psi$ of the following form:
\begin{equation}
\delta \psi = A \exp\left\{ -\frac{1}{2} \frac{y^2 + z^2}{\left[\sigma_\perp \left(1 - \epsilon \cos(k x)\right)\right]^2} \right\},
\end{equation}
where $\sigma_\perp$ denotes the radial width of the stationary state, $\epsilon \ll 1$, and the modulation wave vector is given by $k = 2\pi/\lambda = 2\pi/L_{}$, ensuring that a single modulation wavelength fits within the longitudinal size $L$ of the computational box.
With this prescription, the initial states of the two components read $\psi_i = \psi_i^{\mathrm{st}} + \delta \psi$,
and are subsequently evolved according to the GP equations \eqref{eq:gpe} \footnote{The time-dependent GP equations are solved using a split-step Fourier method; see, e.g., Ref.~\cite{jackson1998}.}. 

\begin{figure*}
\includegraphics[width = 1.8\columnwidth]{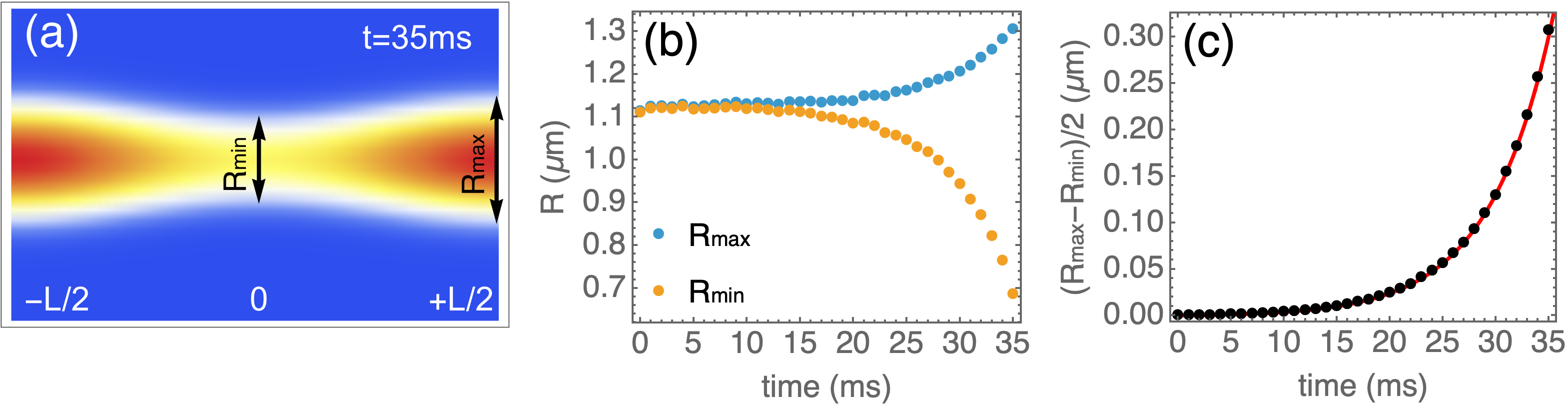}
\caption{
Results of the GP simulations for
 the case $N/L=2.1\times10^4 \mu\mathrm{m}^{-1}$, $L_{}=12.5 \mu\mathrm{m}$. 
(a) Total density profile of the filament, integrated along the transverse $z$ direction, at $t=35$ ms, after the onset of the instability. 
The arrows indicate the positions at which the radial widths $R_{\mathrm{min}}$ and $R_{\mathrm{max}}$ are computed. 
(b) Time evolution of $R_{\mathrm{min}}$ and $R_{\mathrm{max}}$. 
(c) Time evolution of the quantity $(R_{\mathrm{max}}-R_{\mathrm{min}})/2$ (black data points) along with the exponential fit from Eq.~\eqref{fitfunc} (red solid line). 
}
\label{fig:fig4}
\end{figure*}

If $k$ corresponds to the wavevector of an unstable mode, a neck spontaneously forms in the middle of the filament during the evolution, as shown in Fig.~\ref{fig:fig4}(a) for a representative case.
To characterize the growth of the unstable modes, we follow the approach of Ref.~\cite{Ancilotto2023}, measuring the time evolution of the radial size--defined as the value of $r_{\perp}$ at which the total density drops to half of its maximum central value \footnote{Using a different definition of the radial size does not affect the results.}--at two distinct axial positions:  $x = 0$, where we define $R_{\mathrm{min}}$, and $x = L/2$, where we define $R_{\mathrm{max}}$. 
Their corresponding time evolutions are shown in Fig.~\ref{fig:fig4}(b).

Then, by assuming the following time dependence,
\begin{align}
R_{\mathrm{max}}(t) &= R_0 \bigl(1 + \delta e^{\omega t}\bigr), \\
R_{\mathrm{min}}(t) &= R_0 \bigl(1 - \delta e^{\omega t}\bigr),
\end{align}
where $R_0$ is the filament radial size without perturbation, $\delta$ is the perturbation amplitude, and $\omega$ is the growth rate, we can write
\begin{equation}
\left(R_{\mathrm{max}} - R_{\mathrm{min}}\right)/2 = \delta e^{\omega t}.
\label{fitfunc}
\end{equation}
We extract the growth rate $\omega$ by fitting the time evolution of the quantity $(R_{\mathrm{max}} - R_{\mathrm{min}})/2$, obtained from the GP simulations, using the above ansatz \eqref{fitfunc}, as illustrated in Fig.~\ref{fig:fig4}(c). 
By repeating this analysis for different computational box lengths $L$, and thereby varying the modulation wavevector $k = 2\pi/L$, we reconstruct the dispersion relation $\omega(k)$ of the unstable modes.
This is shown in Fig.~\ref{fig:fig3}(a) for the three values of the linear density $N /L$, where data extracted from the GP simulations (data points) are compared with the results of the BdG analysis based on the effective single-component approximation discussed earlier (dashed lines).
The excellent agreement confirms that the single-component BdG approach accurately reproduces the results of the full two-component GP simulations. 
Given its significantly lower computational cost, we adopt the single-component BdG method for the analysis presented in the following sections.

\subsection{Interpretation in terms of capillary instability}
\label{sec:RPinterpretation}

We now show that the spectra of the unstable modes presented above exhibit the characteristic features of the capillary instability.
In an infinite filament of an inviscid and incompressible classical liquid, with radius $R$, the growth rate of the unstable modes is given by the Rayleigh--Plateau (RP) dispersion relation:
\begin{equation}
\omega_{RP}^2(k)= \frac{1}{\tau_c^2}\left[ \frac{I_1(k R)}{I_0(k R)}k R\,(1-k^2 R^2)\right],
\label{omkCF}
\end{equation}
where $I_0(x)$ and $I_1(x)\equiv I_0^\prime (x)$
are modified Bessel functions of the first kind, 
and $\tau_c=\sqrt{n R^3/\sigma}$ is the capillary time characterizing the growth rate, which depends on the liquid density $n$ and its surface tension $\sigma$. 
From Eq.~\eqref{omkCF}, it follows that
the spectrum is defined within the range $kR<1$, and is fully determined once the filament radius $R$ and the capillary time $\tau_c$ are given. A universal spectrum for the unstable modes is thus obtained by introducing 
 rescaled quantities, $\tilde{\omega}\equiv \omega_{RP} \tau_c$ and $\tilde{k}\equiv kR$:
\begin{equation}
\tilde{\omega}(\tilde{k}) = \sqrt{ \frac{I_1(\tilde{k})}{I_0(\tilde{k})}\tilde{k}\,(1-\tilde{k}^2)}\,.
\label{omkCF_adim}
\end{equation}
The maximum growth rate, corresponding to the fastest-growing mode
destabilizing the filament, occurs at $\tilde{k}_{\mathrm{max}}=0.697$, with $\tilde{\omega}_{\mathrm{max}}$=0.343. 

To demonstrate the connection between the calculated BdG spectrum $\omega(k) $ and the phenomenon of capillary instability, we fit our numerical data using the Rayleigh--Plateau dispersion relation [Eq.~\eqref{omkCF}]. 
From the fit, we extract the parameters $R$ and $\tau_c$, which are then used to rescale the data. The rescaled spectra, shown in Fig.~\ref{fig:fig3}(b) for the three different values of the linear density $N/L$, are plotted together with the Rayleigh--Plateau prediction (solid line).  All data collapse onto the classical Rayleigh--Plateau curve with excellent agreement.

The fitted parameters $R$ and $\tau_c$ can be compared, respectively, with the radial density profiles obtained from the GP simulations and with an independent estimate of the capillary time based on the expression for a binary mixture reported in Ref.~\cite{Ancilotto2023}:
\begin{equation}
\tau_c = \sqrt{\frac{(m_1\rho_1+m_2\rho_2) R^3}{\sigma }} \, .
\label{tau_cap}
\end{equation}
In the above expression, the surface tension $\sigma$ 
is estimated from the quantum kinetic energy associated with the spatial variation of the calculated density profiles \cite{Ao1998,Jimbo2021}:
\begin{equation}
\sigma =\alpha\sum_{i=1,2} \int\limits_{0}^{+\infty} \frac{\hbar^2}{2 m_i} \left( \frac{d}{dr_{\perp}} \sqrt{|\psi_i(r_{\perp})|^2} \right)^2 \!\!dr_{\perp},
\label{surface_tension}
\end{equation}
where the prefactor is set to $\alpha=2$ consistently with the values reported in Ref.~\cite{Cikojevic2021}, including a correction, in terms of the so-called Tolman length, accounting for the curvature of the surface in our geometry \footnote{It is worth noting  that the variations of the surface tension with the filament's linear density, as well as those arising from the harmonic radial confinement (see Sec.\ref{sec:confinedfilament}), are found to be negligible within the investigated parameter range}.

\begin{figure}[ht]
\begin{center}
\includegraphics[width= 0.9\columnwidth]{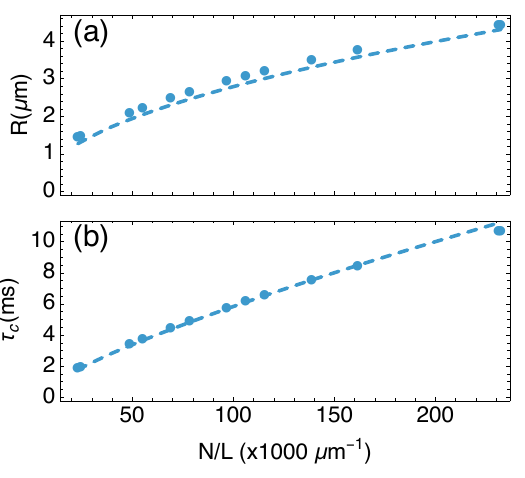}
\caption{(a) Comparison between the fitted value of $R$ (data points) and the radial size of the filament obtained from GP simulations (dashed line), defined as the radius where the density drops to half its peak value, plotted as a function of the linear density. (b) Comparison between the capillary time $\tau_c$ obtained from the fit (data points), with the theoretical prediction from Eq.~\eqref{tau_cap}, where $R$ is taken as the radial size given by the GP simulations, plotted as a function of the linear density.
}
\label{fig:fig5}
\end{center}
\end{figure}

In Fig.~\ref{fig:fig5}(a), we compare the values of $R$ extracted from the fit of the BdG spectra (data points) with the radial size of the \textit{unperturbed filament}  obtained from the GP simulations (dashed line). 
Both quantities increase with the linear density and exhibit the same overall trend.
In Fig.~\ref{fig:fig5}(b), we show the two independent estimates of the capillary time discussed above: one obtained by fitting the unstable mode spectra with the Rayleigh--Plateau formula (data points), and the other calculated from Eq.~\eqref{tau_cap} using the radial size extracted from the GP simulations as input for $R$ (dashed line).

The excellent agreement confirms that the filament’s unstable mode spectrum is fully consistent with capillary instability. 
Remarkably, the dynamics of quantum liquid filaments, such as those studied here, closely mirror the classical Rayleigh--Plateau instability. This analogy holds not only for flat-top filaments (as illustrated in Fig.~\ref{fig:fig2} for $N/L = 20.4 \times 10^4\, \mu\mathrm{m}^{-1}$), where a distinct bulk region can be clearly distinguished, but also for lower linear densities (e.g., $N/L = 2.1\times 10^4\, \mu\mathrm{m}^{-1}$), where identifying the filament surface and radius becomes more challenging.

\section{Radially confined filament}
\label{sec:confinedfilament}

\begin{figure}
\begin{center}
\includegraphics[width=1.0\linewidth,clip]{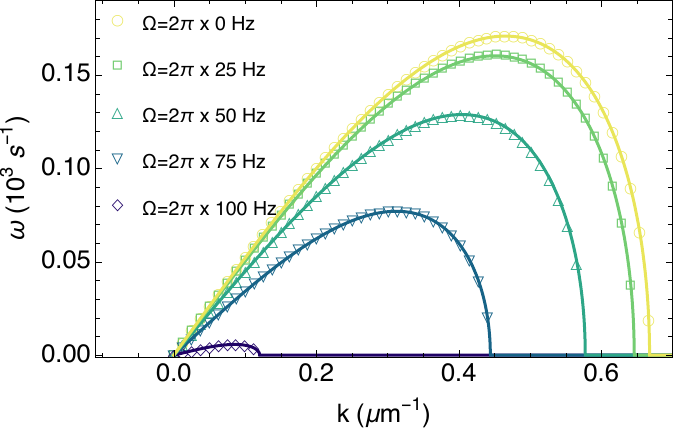}
\end{center}
\caption{Spectra of unstable modes for a filament with linear density $N/L=2.1\times10^4 \mu\mathrm{m}^{-1}$, confined in a harmonic radial waveguide. Results are shown for different values of the radial trap frequency $\Omega$ (see legend). 
The solid lines rare fits using Eq.~\eqref{omkCF_guide}, with $R$ and $\tau_c$ as free parameters. The spectra are computed using the single-component BdG approach. }
\label{fig:fig6}
\end{figure}

We now turn to the effect of radial confinement, a situation that naturally arises in atomic condensates confined in waveguide-like geometries. In analogy with classical liquids in a pipe — where hard-wall boundaries suppress fragmentation (see, e.g., Ref.~\cite{Patrascu2019}) — we expect that strong radial confinement can inhibit the onset of the instability. In the following, we focus on harmonic radial confinement, which is both experimentally relevant and straightforward to implement in ultracold atom setups.
As before, we compute the growth rate $\omega(k)$ of the unstable modes at fixed linear density $N/L$, considering a  filament confined by the potential defined in Eq.~\eqref{eq:potential}. 
The resulting spectra, shown in Fig.~\ref{fig:fig6} for different values of the trap frequency $\Omega$,  exhibit a marked dependence on the radial confinement. As $\Omega$ increases, the instability region becomes progressively narrower and eventually vanishes, indicating the stabilization of the trapped filament.

\begin{figure}
\begin{center}
\includegraphics[width=0.9\linewidth,clip]{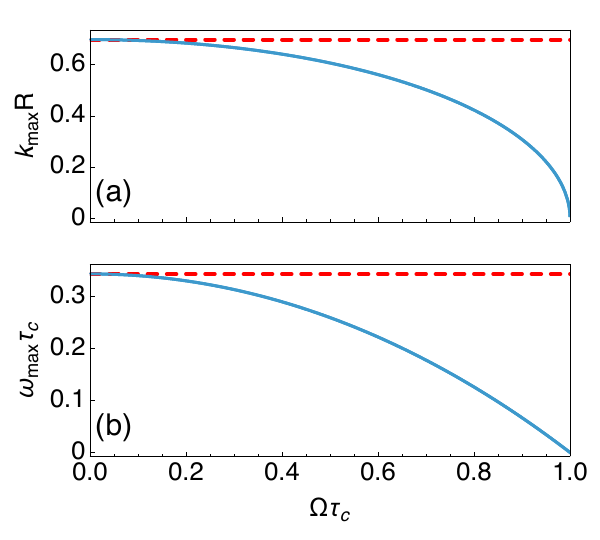}
\end{center}
\caption{Behavior of the rescaled quantities $k_{max}R$ and $\omega_{\mathrm{max}}\tau_c$ as functions of the transverse frequency $\Omega$ for the trapped filament, as obtained from Eq.~\ref{omkCF_guide}.
The red dashed lines indicate the corresponding values for a free filament.}
\label{fig:fig7}
\end{figure}

Interestingly, radial confinement, leads to a slight decrease in the filament radius $R$, of the order of few percent within the investigated range.
Consequently the capillary time, which scales as $\sqrt{R}$, also decreases. According to the Rayleigh--Plateau scaling law given in Eq.~\eqref{omkCF}, one would thus expect both the most unstable wavevector, $k_{\mathrm{max}} = 0.697/R$, and the corresponding frequency, $\omega_{\mathrm{max}} = 0.343/\tau_c$, to increase with stronger radial confinement. However, this expectation of enhanced instability is in stark contrast with the behavior observed in Fig.~\ref{fig:fig6}. 
This apparent puzzle is resolved by the fact that, as recently shown in Ref.~\cite{Patankar2022} for an inviscid liquid cylinder, the dynamical instability of the system is altered by the presence of a radial force. 
By adapting the results of Ref.~\cite{Patankar2022} [see Eq.~(2.3) therein] to our case of static harmonic confinement—corresponding to a radial force per unit mass $\mathcal{F} = -\Omega^2 r$—the spectrum of unstable axisymmetric modes becomes
\begin{equation}
\omega_{RP}^{\mathrm{conf}}\!(k)= \frac{1}{\tau_c}\sqrt{\frac{I_1(kR)}{I_0(kR)}\left[ kR\,(1-(kR)^2-(\Omega\tau_c)^2) \right]}\,,
\label{omkCF_guide}
\end{equation}
providing a generalization of the free-space scaling in Eq.~\eqref{omkCF} to the presence of radial confinement.
This result shows that the unstable region shrinks progressively as
\begin{equation}
kR < \sqrt{1 - (\Omega \tau_c)^2},
\end{equation}
and eventually vanishes at the \textit{critical frequency}
\begin{equation}
\Omega_{c} = \tau_c^{-1}.
\end{equation}
Since $\tau_c$ is expected to increase with the linear density $N/L$, the corresponding critical frequency 
$\Omega_{c}$ required for stabilization is expected to decrease accordingly. 
The overall behavior of the rescaled quantities $k_{max}R$ and $\omega_{\mathrm{max}}\tau_c$ as functions of the transverse frequency $\Omega$ is shown in Fig.~\ref{fig:fig7}, together with their corresponding values for a free filament (red dashed lines). 
Both $k_{\mathrm{max}}$ and $\omega_{\mathrm{max}}$ tend to vanish with increasing transverse frequency, effectively suppressing the capillary instability.

\begin{figure}[ht]
\begin{center}
\includegraphics[width= 0.9\columnwidth]{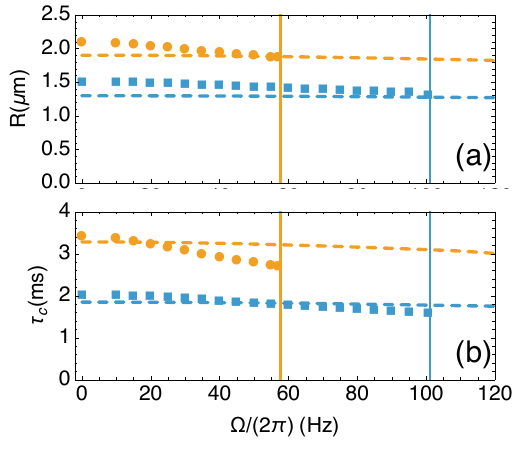}
\caption{(a) Comparison between the fit parameter $R$ (data points) and the radial size of the filament (dashed line) obtained from GP simulations. (b) Comparison between the capillary time obtained from the fit (data points), with the theoretical one from Eq.~\eqref{tau_cap}, where $R$ is taken as the radial size from GP simulations. In both panels the parameters are plotted as a functions of the 
transverse frequency $\Omega$. Yellow circles) and blue squares correspond to $N/L=2.1\times10^4$ and  $4.2\times 10^4 \mu\mathrm{m}^{-1}$ respectively. Vertical lines indicate the critical frequency beyond which the filament becomes stable. 
}
\label{fig:fig8}
\end{center}
\end{figure}

Similarly to the discussion in the previous section, the spectra of the unstable modes of the trapped filament can be fitted using Eq.~\eqref{omkCF_guide} to extract the parameters $R$ and $\tau_c$, which can then be compared with the corresponding estimates obtained from the calculated density profiles. This comparison is shown in Fig.~\ref{fig:fig8} for two different values of the linear densities, as a function of the radial trapping frequency $\Omega$. In estimating $\tau_c$, we use a surface tension value derived from Eq.~\eqref{surface_tension},  where $\psi _i(r)$ are the stationary-state wavefunctions obtained in the presence of the radial confinement.
The agreement between the values of the radius and the capillary time obtained from the calculated density profiles (dashed lines) and those extracted from the fit (data points) is overall satisfactory. Although the fitting expression in Eq.~\eqref{omkCF_guide} was originally derived for an incompressible liquid filament with uniform density \cite{Patankar2022}, it still captures the main trends of both $R$ and $\tau_c$ in the present case. This consistency highlights the robustness of the underlying hydrodynamic analogy, even in the presence of density inhomogeneities, quantum corrections, and harmonic confinement.
Moreover, the vertical lines in Fig.~\ref{fig:fig8}, which indicate the stabilization frequencies, clearly confirm the expected decrease of the stabilization frequency with increasing linear density.

To conclude this section, we comment on the implications of the above analysis for the experiment by Cavicchioli \textit{et al.} \cite{Cavicchioli2025}. In that experiment, a droplet is allowed to expand in a waveguide, and is observed to stretch into a filament of finite length and eventually break up into several smaller droplets. Remarkably, the observed dynamics have been accurately interpreted in terms of the Rayleigh--Plateau formula \eqref{omkCF}, without explicitly accounting for the transverse confinement imposed by the waveguide. This apparent discrepancy with the present findings can be resolved by considering the nature of the filament in the experiment, which originates from a droplet excited in an elongation-compression mode and then allowed to freely expand along the axial direction, without any constraints. This scenario clear contrasts with the stationary configurations considered here, where the system is constrained to have a fixed linear density, and translational invariance suppresses axial dynamics. In this case, radial confinement directly affects the only available degree of freedom for accommodating perturbations and can thus stabilize the filament against capillary instability.

\section{Conclusions}
\label{sec_conclusions}
\label{conclusions}

In conclusion, we have investigated the spectrum of unstable modes of a quantum filament composed of a heteronuclear atomic mixture in the self-bound, liquid-like regime.
The spectrum has been obtained using the Bogoliubov--de Gennes linear stability analysis within a single-component effective functional framework, whose validity has been confirmed by comparison with full numerical simulations of the coupled extended Gross--Pitaevskii equations for the two atomic species.

Our results reveal a robust and quantitative parallel between the unstable-mode spectra of quantum filaments—both in free space and under harmonic radial confinement—and the classical capillary instability of an inviscid liquid filament. 
For a free-space filament, the spectrum perfectly matches the Rayleigh--Plateau prediction. In the presence of radial confinement, the instability is progressively suppressed, and eventually vanishes beyond a critical frequency, set by the inverse of the capillary time and dependent on the linear density of the filament.
This behavior is accurately captured by adapting a classical hydrodynamic model \cite{Patankar2022}, originally developed for oscillatory forcing, to the case of static harmonic confinement.
The similarities and differences with the analysis performed for the experiment by Cavicchioli \textit{et al.} \cite{Cavicchioli2025} have also been briefly discussed.

These findings further highlight the potential of ultracold quantum gases as model systems for investigating hydrodynamic instabilities in regimes where quantum effects and confinement play a crucial role.
The agreement with classical analogue not only reinforces the hydrodynamic interpretation but also opens avenues for exploring more complex phenomena, such as interfacial effects in multicomponent systems or nonlinear dynamics beyond the initial growth stage.
An interesting direction for future work would be to explore alternative confinement geometries or time-dependent confinements, to test the generality and limitations of the suppression mechanism. In particular, the ring geometry may enable the implementation of a real system which, in the connected geometry, reproduces the physics of an infinitely long filament.

\begin{acknowledgments}
We thank A.~Burchianti, and L.~Cavicchioli,  for valuable discussion and for their critical reading of the manuscript. 
This work was supported by the European Union - NextGeneration EU, within PRIN 2022, PNRR M4C2, Project QUANTAMI 20227JNCWW (CUP B53D23005010006), 
PNRR MUR project PE0000023-NQSTI and the ‘Integrated infrastructure initiative in Photonic and Quantum Sciences’ I-PHOQS (IR0000016, ID D2B8D520, CUP B53C22001750006). M.M. acknowledges support from Grant No.~PID2021-126273NB-I00 funded by MCIN/AEI/10.13039/501100011033 and by “ERDF A way of making Europe”, as well as from the Basque Government through Grant No. IT1470-22.
\end{acknowledgments}

\appendix

\section{BdG equations}
\label{sec:appendix}

In order to study the elementary excitations of the system, we expand the wave function in the form 
\begin{equation}
\Psi(\bm{r},t)=e^{-i\mu{t}/\hbar}[\Psi _0(\bm{r})+
u(\bm{r})e^{-i\omega t}-v^*(\bm{r})e^{i\omega t}].
\end{equation}
Substituting this expression into the time-dependent GP equation [see Eqs.~\eqref{eq:acca}, \eqref{eq:nlgpe}],
\begin{equation}
i \hbar \partial_{t}\Psi(\bm{r}) 
=[-\alpha \nabla ^2+U(\bm{r})+V^{\mathrm{ext}}(\bm{r})]\Psi(\bm{r}), 
\label{eq:tdschrod}
\end{equation} 
with 
\begin{equation}
U(\bm{r}) = 2\beta \rho(\bm{r}) + \frac {5}{2}\gamma \rho^{3/2}(\bm{r}) 
= U^{MF}(\bm{r})+U^{QF}(\bm{r}),
\end{equation}
where $\rho=|\Psi|^{2}$, and keeping only terms linear in the functions $u$ and $v$, one obtains the following coupled equations:
\begin{widetext}
\begin{align}
\label{bdgeq}
\hbar \omega u(\bm{r}) & =
 [-\alpha \nabla ^2+V^{\mathrm{ext}}(\bm{r})-\mu ]u(\bm{r}) +4\beta \Psi _0^2(\bm{r}) u(\bm{r}) -2\beta \Psi _0^2(\bm{r}) v(\bm{r}) 
+ 
\frac{25}{4}\gamma \Psi _0^3(\bm{r}) u(\bm{r})
- 
\frac{15}{4}\gamma \Psi _0^3(\bm{r}) v(\bm{r}),
\\
-\hbar \omega v(\bm{r}) & =
 [-\alpha \nabla ^2+V^{\mathrm{ext}}(\bm{r})-\mu ]v(\bm{r}) +4\beta \Psi_0 ^2(\bm{r}) v(\bm{r}) -2\beta \Psi _0^2(\bm{r}) u(\bm{r}) 
+ 
\frac{25}{4}\gamma \Psi_0^3(\bm{r}) v(\bm{r}) 
- 
\frac{15}{4}\gamma \Psi_0 ^3(\bm{r}) u(\bm{r}).
\end{align}
\end{widetext}

To study the excitations of an infinitely extended liquid filament, we perform numerical calculations on a finite-sized system confined within an orthorhombic cell of dimensions $L_x$, $L_y$, and $L_z$, with periodic boundary conditions (a \textit{periodic supercell}).
To this end we expand the (real) function $\Psi _0(\bm{r})$ and the complex functions $u(\bm{r}),v(\bm{r})$ in the Bloch form appropriate to a periodic system:
\begin{align}
\Psi_0(\bm{r}) &= \sum_{\bm{G}} \Phi_{\bm{G}} e^{i\bm{G} \cdot \bm{r}} 
\\
u_{n,\bm{k}}(\bm{r}) &= e^{i\bm{k} \cdot \bm{r}}\sum_{\bm{G}} 
u_{\bm{k} + \bm{G}}^{(n)} e^{i\bm{G} \cdot \bm{r}}, 
\label{bloch}
\end{align}
and an analogous expression for $v_{n,\bm{k}}(\bm{r})$.

In the above expansions, the $\bm{G}$-vectors are the reciprocal lattice vectors associated with the periodic supercell used for the real-space calculations, and the wavevector $\bm{k}$ belongs to the first Brillouin zone.
By making the above substitutions in Eq.~\eqref{bdgeq} (and omitting the band index $n$ for simplicity) one gets:
\begin{widetext}
\begin{align}
\label{bdgeq_g}
\hbar \omega u_{\bm{k} +\bm{G}} &= \left[\alpha (\bm{k}+\bm{G})^2-\mu \right]u_{\bm{k} +\bm{G}} 
+ \sum_{\bm{G}^\prime}\left(
V^{\mathrm{ext}}_{\bm{G}-\bm{G}^\prime}
+ 2\tilde {U}^{MF}_{\bm{G}-\bm{G}^\prime}
+ \frac {5}{2} \tilde {U}^{QF}_{\bm{G}-\bm{G}^\prime}
\right) u_{\bm{k}+\bm{G}^\prime}
-\sum_{\bm{G}^\prime} \left(
\tilde {U}^{MF}_{\bm{G}-\bm{G}^\prime}
+ \frac {3}{2} \tilde {U}^{QF}_{\bm{G}-\bm{G}^\prime} 
\right) v_{\bm{k}+\bm{G}^\prime}, 
\\
\hbar \omega v_{\bm{k}+\bm{G}} &= -\left[{\alpha}(\bm{k}+\bm{G})^2-\mu \right]v_{\bm{k} +\bm{G}}
-\sum_{\bm{G}^\prime} \left( 
V^{\mathrm{ext}}_{\bm{G}-\bm{G}^\prime} 
+ 2\tilde {U}^{MF}_{\bm{G}-\bm{G}^\prime}
+ \frac {5}{2} \tilde {U}^{QF}_{\bm{G}-\bm{G}^\prime} 
\right)v_{\bm{k}+\bm{G}^\prime}
+\sum_{\bm{G}^\prime} \left(\tilde {U}^{MF}_{\bm{G}-\bm{G}^\prime}
+\frac {3}{2} \tilde {U}^{QF}_{\bm{G}-\bm{G}^\prime} \right)u_{\bm{k}+\bm{G}^\prime} \nonumber.
\end{align}

\end{widetext}
Here $\tilde{U}^{MF}_{\bm{G}}$, $\tilde{U}^{QF}_{\bm{G}}$ and $V^{\mathrm{ext}}_{\bm{G}-\bm{G}^\prime} $
are the Fourier components, respectively, of the mean-field interaction potential $U^{MF}(\bm{r})$, 
of the potential term $U^{QF}(\bm{r})= ({5}/{2})\gamma \rho ^{3/2}$ 
accounting for quantum fluctuations within the LHY approximation, 
and of the external potential $V^{\mathrm{ext}}(\bm{r})$:

\begin{widetext}
\begin{equation}
U^{MF}(\bm{r})=
\sum _{\bm{G}} \tilde {U}^{MF}_{\bm{G}} e^{i\bm{G}\cdot \bm{r}}, \quad
U^{QF}(\bm{r})=
\sum _{\bm{G}} \tilde {U}^{QF}_{\bm{G}} e^{i\bm{G}\cdot \bm{r}}, \quad 
V^{\mathrm{ext}}(\bm{r})=
\sum _{\bm{G}} \tilde V^{\mathrm{ext}}_{\bm{G}} e^{i\bm{G}\cdot \bm{r}}.
\label{eq:utilde}
\end{equation}
\end{widetext}

The above system of equations can be conveniently recast in a more compact form by defining the matrices $A$ and $B$ [with dimensions $(n_x \, n_y \, n_z)^3\times (n_x \, n_y \, n_z)^3$, where $\{n_x,n_y,n_z\}$ is the real-space mesh used to integrate the stationary equation \eqref{eq:acca}], defined in Eq.~\eqref{mat_A} and Eq.~\eqref{mat_B} in the main text.

\bibliography{bibliography}

\end{document}